# NOMAD 2018 Kaggle Competition: Solving Materials Science Challenges Through Crowd Sourcing

Christopher Sutton,[1,*] Luca M. Ghiringhelli,[1,*]
Takenori Yamamoto,[2] Yury Lysogorskiy,[3] Lars Blumenthal,[4,5]
Thomas Hammerschmidt,[3] Jacek Golebiowski,[4,5] Xiangyue Liu,[1] Angelo Ziletti,[1]
Matthias Scheffler[1]

[1]*Fritz Haber Institute of the Max Planck Society*
*Berlin, Germany*

[2] *Research Institute for Mathematical and Computational Sciences (RIMCS), LLC*
*Yokohama, Japan*

[3] *ICAMS*
*Ruhr-Universität*
*Bochum, Germany*

[4] *EPSRC Centre for Doctoral Training on Theory and Simulation of Materials*
*Department of Physics*
*Imperial College London*
*London, U.K.*

[5] *Thomas Young Centre for Theory and Simulation of Materials*
*Department of Materials*
*Imperial College London*
*London, U.K.*

[*]Corresponding authors: sutton@fhi-berlin.mpg.de; ghiringhelli@fhi-berlin.mpg.de

# Abstract

Machine learning (ML) is increasingly used in the field of materials science, where statistical estimates of computed properties are employed to rapidly examine the chemical space for new compounds. However, a systematic comparison of several ML models for this domain has been hindered by the scarcity of appropriate datasets of materials properties, as well as the lack of thorough benchmarking studies. To address this, a public data-analytics competition was organized by the Novel Materials Discovery (NOMAD) Centre of Excellence and hosted by the on-line platform Kaggle using a dataset of 3,000 $(Al_xGa_yIn_z)_2O_3$ compounds (with $x+y+z = 1$). The aim of this challenge was to identify the best ML model for the prediction of two key physical properties that are relevant for optoelectronic applications: the electronic band gap energy and the crystalline formation energy. In this contribution, we present a summary of the top three ML approaches of the competition including the 1st place solution based on a crystal graph representation that is new for ML of the properties of materials. The 2nd place model of this competition combined many candidate descriptors from a set of compositional, atomic environment-based, and average structural properties with the light gradient-boosting machine regression model. The 3rd place model employed the smooth overlap of atomic positions representation with Gaussian process regression. To gain insight into whether the representation or the regression model determines the overall model performance, nine ML models comprised of the top-three representations from the competition and their regression models were examined using the Pearson correlation among their prediction errors. At a fixed representation, the largest correlation is observed in predictions made with kernel ridge regression (or Gaussian process regression) and a neural network, reflecting a similar performance on the same test set samples. Averaging the two models with the smallest Pearson correlation yields an even higher prediction accuracy.

# Introduction

Computational approaches have become a powerful tool for guided design of new compounds to potentially aid the development of advanced technologies. However, the identification and discovery of new materials that are ideal for targeted applications is a nontrivial task that requires examining enormous compositional and configurational degrees of freedom. For example, an alloy with two substitutional atoms in the unit cell and with $M$ sites displays a large number of possible configurational states of the order of $2^M$ (neglecting symmetry) for each lattice, and most often several polymorphs have to be examined.

Density-functional theory (DFT) typically provides the best compromise between accuracy and cost; nevertheless, a single energy evaluation using DFT scaling as a high-order polynomial with system size. As a result of the high computational demand, DFT-based exploration of configurational spaces of alloys is only feasible for unit cells with a relatively small number of atoms. To efficiently search this vast chemical space, methods that allow for fast and accurate estimates of materials properties have to be developed.

Machine learning (ML) promises to accelerate the discovery of novel materials by allowing to rapidly screen candidate compounds at significantly lower computational cost than traditional electronic structure approaches.[1-6] A key consideration for an ML model of material properties is how to include atomic and structural information as a fixed-length feature vector to enable regression, which is referred to as the representation or descriptor. Given that knowledge of the atomic positions and chemical species (e.g., the atomic number) for a given system is sufficient to construct the Hamiltonian, a ML descriptor should include the geometrical and chemical information in a convenient way. A considerable amount of work has been devoted to defining suitable ML descriptors of molecules or materials by encoding the chemical and geometrical information in various ways such as Coulomb matrices,[7,8] scattering transforms,[9] diffraction patterns,[10] bags of bonds,[11] many-body tensor representation,[12] smooth overlap of atomic positions (SOAP),[13,14] and several symmetry-invariant transformations of atomic coordinates.[15-17]

All of these approaches represent the training or test samples and are typically combined with kernel ridge regression (KRR) or Gaussian process regression (GPR)[18] methods to effectively identify differences in the structures of the data set. In addition, generalized atom-centered symmetry functions have also been developed to be combined with a neural network (NN).[19,20] Other approaches such as a modified Least Absolute Shrinkage and Selection Operator (LASSO)[21] and the Sure Independence Screening and Sparsifying Operator (SISSO)[22] have focused on identifying the best descriptor out of a large space of mathematical combinations of simple features that represent the chemical information and (currently only simplified) structural information.[23-25]

Of particular importance for the efficient modeling of the large configurational space of substitutional alloys, the cluster expansion (CE) method[26-32] is an ML representation using only an occupational variable for each substitutional lattice site. However, the lack of explicit local atomic information (e.g., bond distances and angles) of the crystalline systems prevents a broad and transferable application of this approach. Along these same lines, semiempirical interatomic potentials or force field-based approaches use parameterized models based on classical mechanics (e.g., short range two-body and three body interactions, long range-Coulomb interactions) to approximate quantum mechanical properties.

With so many choices of the various structural representations, it is often unclear which will be the most insightful or accurate for a given problem. Furthermore, optimizing an ML model for a particular application can be a time-consuming endeavor: A given representation is combined with a specific regression model (i.e., a model class and an induction algorithm) whose hyperparameters are tuned subsequently. Therefore, typically, only a few combinations of representation and regression algorithms are carefully tested for a specific application, which limits the understanding of how well various ML models perform. Crowd sourcing offers an alternative approach for examining several ML models by identifying a key problem and challenging the community to solve it by proposing solutions that are ranked in an unbiased way. To this end, the Novel Materials Discovery (NOMAD)[33] Centre of Excellence organized a data-analytics competition for predicting the key properties of transparent conducting oxides (TCOs) with Kaggle, which is one of the most recognized online platforms specializing in hosting data-science competitions.

TCOs are an important class of well-developed and commercialized wide band-gap materials that have been employed in a variety of (opto)electronic devices such as solar cells, light-emitting diodes, field-effect transistors, touch screens, sensors, and lasers.[34-44] However, only a small number of compounds display both transparency and electronic conductivity suitable enough for these applications. For example, tin-doped indium oxide ($In_2O_3$:Sn) serves as the primary transparent electrode material for (opto)electronic devices because of its high-transparency over the visible range, resulting from an electronic band gap energy of 2.7 eV,[45,46] and its high electrical conductivity,[47-49] which are typically competing properties. A wide range of experimental band gap energies from 3.6 to 7.5 eV have been reported from alloying $In_2O_3$/$Ga_2O_3$ or $Ga_2O_3$/$Al_2O_3$,[50-56] which suggest that alloying of group-III oxides is a viable strategy for designing new wide band gap semiconductors. However, $Al_2O_3$, $Ga_2O_3$, and $In_2O_3$ all display very different ground-state structures. Therefore, it is unclear which structure will be stable for various compositions. The goal of the competition was to identify the best ML model for both the formation energy (an indication of the stability) and the band gap energy (an indication of transparency) using a dataset that contained 3,000 $(Al_xGa_yIn_z)_2O_3$ compounds, 2,400 of which were used for the training set, with the remaining 600 samples were used as the test set that was kept secret for the entire competition.

The competition was launched on December 18, 2017 and ended on February 15, 2018, attracting 883 participants. Figure 1 shows the distribution of the so-called public and private leaderboard scores for all the participants of the competition. The public score was calculated for only 100 fixed samples from the test set in order to quickly assess the performance of the submitted models, with the two target properties of these samples still kept secret. The remaining 500 samples of the test set were used to determine the winner of the competition, which is displayed in the private leaderboard. The scoring metric used in the competition was the root mean square logarithmic error (RMSLE):

$$\text{RMSLE} = \sqrt{\frac{1}{N}\sum_{i=1}^{N}\left(\log\left(\frac{\hat{y}_i + 1}{y_i + 1}\right)\right)^2}$$

where $N$ is the total number of samples. The error is calculated as the log ratio of the predicted target property $\hat{y}_i$ and corresponding reference value $y_i$ of the formation energy and band gap energy computed using DFT with the PBE exchange-correlation functional using the all-electron electronic structure code FHI-aims with tight settings.[57] The error for both of these two target properties is then averaged for a final assessment of the model performance. The log ratio of the errors is a convenient choice because it prevents the band gap, which is an order of magnitude larger than the formation energy (see Figure S1), from dominating an analysis of the predictive capability of each model.

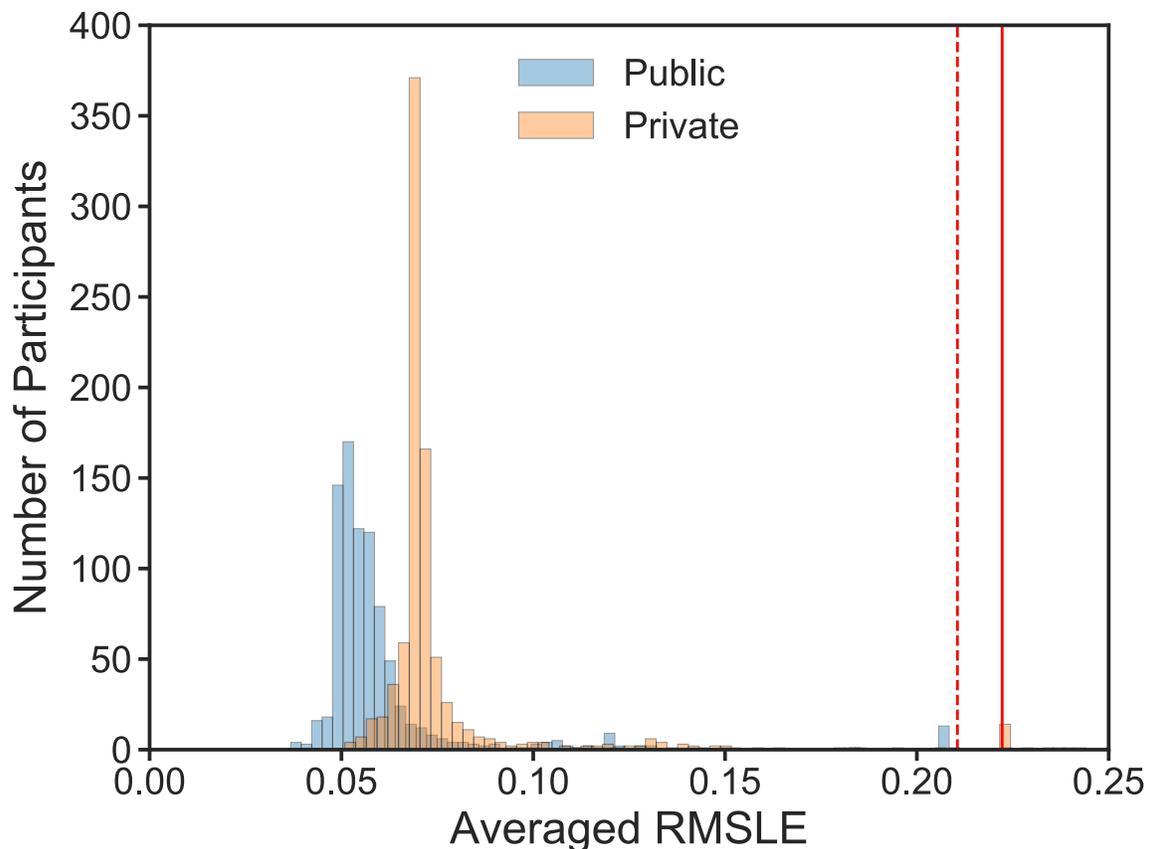

Figure 1. Histogram of averaged RMSLE of the band gap and formation energies for all of the 883 models submitted in the NOMAD 2018 Kaggle competition. The scores are shown for the Kaggle public and private scoreboards of the test containing 600 samples with the values of these two target properties withheld for the entire competition. The public score was calculated for 100 fixed samples; the private score was calculated for 500 samples and was used to determine the winner of the competition. The vertical red lines correspond to the predictions from taking the average value of the training set to predict the public (dashed line) and private (line) datasets.

For the practical application of ML models for high-throughput screening, it is of particular importance to have a model that inputs structural features based on a generalized unrelaxed geometry because the relaxed structures are not readily available. If the relaxed geometry needed to be calculated to obtain the input features for an ML model, then all of the quantities of interest would already be available. In this competition, the structures were provided by the linear combination of the stoichiometric amounts of the $Al_2O_3$, $Ga_2O_3$, and $In_2O_3$ geometries at the same lattice symmetry (i.e., obtained by applying Vegard's law[58,59] for the lattice vectors to generate the input structures); however, the target properties provided for learning and testing correspond to the values from the fully relaxed geometries.

In Section I of this contribution, we describe the performance of the three ML approaches on the original dataset provided in the NOMAD 2018 Kaggle competition. Section II provides a comparison in the performance of these three representations with various regression methods to gain an understanding of the key determining factors for the high performance of the winning models. Section III provides a comparison between the errors of the fully optimized geometries and those obtained using the starting structures generated using Vegard's law. Section IV examines the generalization error of the ML models for lattice symmetries outside of the training set. A detailed description of each of the three winning ML models from the competition is provided in the Methods section, we only briefly describe the models in the main text.

# Results

## I. Performance of the three winning approaches from the NOMAD 2018 Kaggle competition

As already mentioned in the introduction, the errors in both the band gap and formation energy of the crystalline system differ by about an order of magnitude in their mean and standard deviations. Thus, simply averaging the two absolute errors would result in an error metric that is dominated by the band gap energy because of its larger magnitude (Figure S1). This is why the RMSLE was the performance metric used in the competition. However, we decided for the discussion in this section to use the mean absolute errors (MAE) of the band gap and formation energies separately because they allow for a more intuitive quantification of model performance from a physical point of view:

$$\text{MAE} = \frac{1}{N}\sum_{i=1}^{N}|\hat{y}_i - y_i|.$$

Table 1 compares the RMSLE and MAE for the top three models. The 1$^{st}$ place model employed a crystal graph representation to convert the crystal structure into features by counting the contiguous sequences of unique atomic sites of various lengths (called *n*-grams), which was combined with kernel ridge regression (KRR).[60] The 2$^{nd}$ place model used an initially large set of candidate features (*i.e.*, weighted chemical properties as well as atomic-environment representations based on analytic bond-order potentials (BOP)[61-64] and basic geometric measures), which is then optimized and combined with the light gradient-boosting machine (LGBM) regression model,[65] which we label as c/BOP+LGBM. The 3$^{rd}$ place solution used the SOAP representation developed by Bartók et al.[13,14] that incorporates information on the local atomic environment through a rotationally integrated overlap of the Gaussian shaped densities centered at the neighbor atoms, which was combined with a three-layer feed-forward NN (SOAP+NN).

The top three models have a test-set MAE for the formation energy within 2 meV/cation, whereas a larger range of 21 meV is observed for the predictions of the band gap energy (Table 1). We note that for all three models, these errors only vary by 2 meV and 13 meV/cation for the formation energy and band gap energy, respectively, when examining five additional re-partition random 80%/20% splits of the entire 3,000 compound dataset (Table S1). Based on the learning curves provided in Figure S2, the formation energy MAE values of all the three methods converge to within 2 meV/cation relative the error when training on the full 2,400 samples for training set sizes ≥ 960. For the band gap energies, a test-set MAE ≤ 16 meV relative to the error obtained when training on the full 2,400 samples is achieved for 960, 1,440, and 1,920 training samples for SOAP+NN, c/BOP+LGBM, and $n$-gram+KRR, respectively.

Overall, the higher accuracy in the formation energy for all three approaches is attributed to the inclusion of the local atomic topography in each model. The lower accuracy for the band gap energy is attributed partly to the fact that the valence band is determined by hybridization of oxygen atoms, whereas the conduction band is described by the metal-metal interactions. Therefore, an accurate description of this property most likely requires additional information to be included in the representation beyond the local structure.

Table 1. A summary of the three winning models of the competition with the test-set root mean square log error (RMSLE) and mean absolute error (MAE) of the formation energy and band gap energy.

| Ranking | ML representation + regression method | Band gap energy | | Formation energy | |
|---|---|---|---|---|---|
| | | Root mean square log error | Mean absolute error (meV) | Room mean square log error | Mean absolute error (meV/cation) |
| 1st | $n$-gram+KRR | 0.077 (0.078*) | 114 (106)* | 0.021 (0.020*) | 15 (14)* |
| 2nd | c/BOP+LGBM | 0.081 | 93 | 0.022 | 15 |
| 3rd | SOAP+NN | 0.081 | 98 | 0.021 | 13 |

*Determined using the quadgram vector instead of the ensemble of trigram and quadgram that was used in the competition.

## II. Three winning representations combined with three regression methods

To understand the effect of the choice of representation vs. regression model on the overall error, we now examine the performance of each representation combined with KRR/GPR, NN, and LGBM. A detailed description of each of the nine models is provided in the Methods section; here we only note that the hyperparameters are optimized for each representation and regression method combination.

The primary goal for training an ML model is accurately generalize the rules learned on the training set to make predictions on unseen data. Overfitting describes the propensity of an ML model to give a higher accuracy on the training set compared with the test set, which is an indication of poor generalizable predictions of the model. To evaluate the generalizable error, we investigate the difference between the 95% percentiles of the MSE for the training and test sets for each of the nine ML models (Δ95%). The 95% percentiles for the training set and test set are given by the upper edges of the boxplots in Figure 2 (The explict values for the MAE and 95% percentiles are provided in Table S2).

Beginning with a discussion of the errors in the formation energy, a practically identical error is observed among the predictions from all the three regression models (KRR/GPR, NN, LGBM) using the c/BOP, SOAP, and $n$-gram representations, with a maximum difference of 4 meV/cation, 2 meV/cation and 3 meV/cation, respectively (Figure 2). However, a large variation of the Δ95% value between the training and test predictions is observed. For example, a consistently larger Δ95% value is calculated when the NN and LGBM regression methods are used irrespective of the three representations. This is apparent in Figure 2 with the much narrower distribution of the errors training set absolute errors (blue) compared to the test set absolute errors (red). More specifically, a markedly large Δ95% is observed for $n$-gram+NN (Δ95% = 36 meV/cation) and $n$-gram+LGBM (Δ95% = 49 meV/cation) compared with $n$-gram+KRR (Δ95% = 20 meV/cation). A similar trend is found for SOAP representation combined with NN (Δ95% = 36 meV/cation), LGBM (Δ95% = 54 meV/cation) and GPR (Δ95% = 31 meV/cation). A slightly larger difference between the 95% confidence thresholds of the training and test sets is computed for c/BOP+LGBM (Δ95% = 39 meV/cation), c/BOP+NN (Δ95% = 28 meV/cation) and c/BOP+KRR (Δ95% = 34 meV/cation). These results indicate a consistently larger Δ95% when the NN and LGBM regression models are used, indicating that these approaches are potentially more prone to overfitting in this application. This observation is consistent with the expectation that overfitting is more likely with highly nonlinear models that have more flexibility when learning a target function. However, this potentially might be resolved by a more careful hyperparameter optimization.

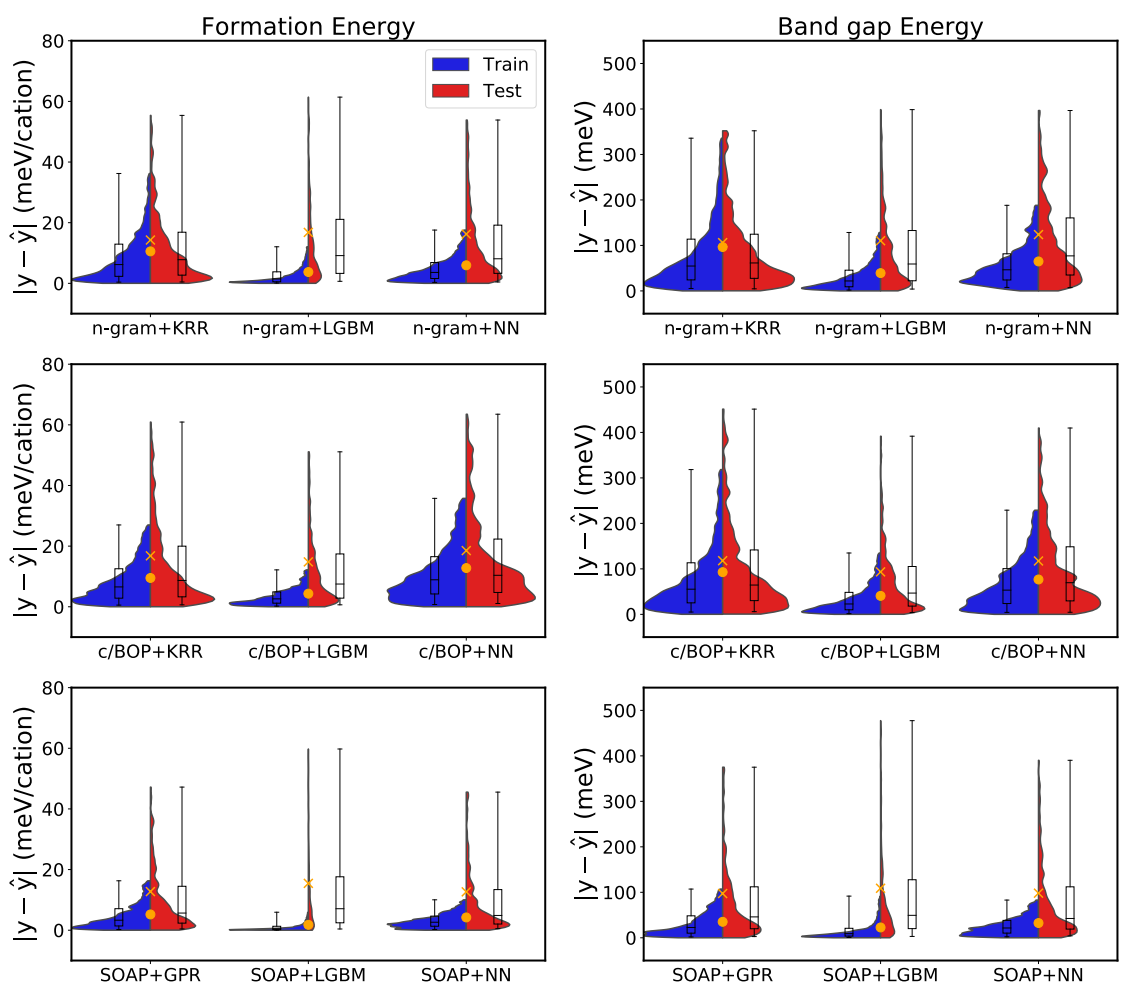

Figure 2. A comparison of the distribution of the absolute errors for the training set (blue) and test set (red) of the formation energy (left) and band gap energy (right) from the three winning representations (*n*-gram, c/BOP, and SOAP) of the competition combined with the KRR/GPR, NN, and LGBM regression models. The markers indicate the mean absolute error (MAE) of the test set (orange cross) and training set (orange filled circle).

Boxplots are included for each training and test set distribution to indicate the 25%, 50%, and 75% percentiles of the absolute errors. The box and violin plots only extend to the 95% percentile. For the training set predictions, the maximum absolute error in the formation (band gap) energy for $n$-gram+KRR, c/BOP+LGBM, and SOAP+NN is 409 meV/cation (1664 meV), 312 meV/cation (1264 meV), and 507 meV/cation (2000 meV), respectively. The corresponding maximum absolute test errors are 289 meV/cation (1083 meV), 276 meV/cation (1680 meV), and 289 meV/cation (1198 meV), respectively.

The Pearson correlation ($r$) between signed errors in the test set predictions is used to quantify correlations between test set errors for all combinations of representation (atomic/BOP, SOAP and $n$-gram) and regression model (LGBM, NN, or KRR/GPR) to elucidate the dominant factors of the model performance through a comparison of the nine model (Figure 3). The Pearson correlation is chosen for this analysis because it is simple parameter-free measure of the linear correlation between two variables (i.e., the residuals between two models) to indicate where two ML models have similar predictions for the test set. Beginning with a discussion of the errors in the formation energy, a practically identical error is observed among the predictions from all the three regression models (KRR/GPR, NN, LGBM) using the c/BOP, SOAP, and $n$-gram representations, with a maximum difference of 4 meV/cation, 2 meV/cation and 3 meV/cation, respectively (Figure 2). The minor variation in the average error is attributed to the dominant effect of the representation in the overall accuracy. However, the range of $r$ values between errors of the three $n$-gram ($r = 0.74 - 0.81$), SOAP ($r = 0.72 - 0.87$), and atomic/BOP ($r = 0.82 - 0.92$) models each using these representations combined with the three different regression model indicates that the accuracy of the three ML models is correlated but not identical. In addition, the highest Pearson correlations in the formation energy errors is observed for the predictions obtained with the c/BOP representation indicating that these models have a strongly correlated description of the test set. Furthermore, among all three representations, the highest Pearson coefficients are consistently obtained for the formation energy residuals between predictions using KRR/GPR and NN, with $n$-gram+KRR vs. $n$-gram+NN ($r = 0.81$), SOAP+GPR vs. SOAP+NN ($r = 0.87$), and c/BOP+KRR vs. c/BOP+NN ($r = 0.92$). In general, the high Pearson correlation among errors of the same representation indicates the choice of the representation is a determining factor in the performance of these approaches.

In contrast to what is observed for the formation energy where the predictions made from the same representations are the most correlated largely independent of the regression model, the bandgap energies are less correlated overall. Relative to what is observed in the errors in the formation energy, a decrease in the $r$ values between the predictions from KRR/GPR and NN regressors for $n$-gram+KRR vs. $n$-gram+NN ($r$ = 0.66), c/BOP+KRR vs. c/BOP+NN ($r$ = 0.73), and SOAP+GPR vs. SOAP+NN ($r$ = 0.80). Overall, these lower Pearson correlation scores for the band gap errors indicate both that even with the same representation, the three respective ML models perform differently for the bandgap predictions, which is potentially a result of the larger errors in this target property.

With an understanding of the correlation for each representation but using different regressors, a key question becomes how correlated the prediction errors are between all nine ML models. The highest correlation is observed when the LGBM regression model is used with the three representations. For the error predictions in the formation energy, the $n$-gram+LGBM vs. SOAP+LGBM ($r$ = 0.78), and c/BOP+LGBM vs. SOAP+LGBM ($r$ = 0.83) show a higher correlation compared with the predictions with $n$-gram+LGBM vs. c/BOP+LGBM ($r$ = 0.74). For the errors in the bandgap energies, the highest correlations in the range of $r$ =0.82 –0.85 are observed between each representation and combined with LGBM, which further demonstrates that this regression model dominates the prediction of this target property. This is rationalized to occur because the LGBM algorithm builds an accurate ML model by ensembling weak learners, which are flowchart-like structures that allow for input data points to be classified based on questions learned from the data.[66] To improve the model predictions, gradient boosting is used to iteratively train additional models on the error. This process specifically addresses the weak points of the previous models, and therefore, the improved correlation indicates that the larger errors become described more consistently by these regression models.

A linear combination of models with uncorrelated errors (i.e., small $r$ values) can perform better than individual ML models, which is the basic idea behind the so-called ensembling.[67-69] To demonstrate that this idea holds for the present data set and set of learners, we have combined various ML models with both small and large Pearson correlations. More specifically, an equivalent error to the 1st place $n$-gram+KRR model (MAE = 14 meV/cation) in the formation energy can be achieved by averaging the predictions from the $n$-gram+NN (MAE = 16 meV/cation) and c/BOP+NN (MAE = 19 meV/cation) models, which have a $r = 0.59$. Furthermore, an MAE = 12 meV/cation can be obtained by averaging the predictions from the 1st place $n$-gram+KRR model (MAE = 14 meV/cation) with SOAP+GPR (MAE = 13 meV/cation), which have an $r = 0.72$. In contrast, ensembling from two models with a large correlation of $r = 0.92$ such as c/BOP+KRR model (MAE = 17 meV/cation) with SOAP+NN (MAE = 13 meV/cation), leads to an MAE = 13 meV/cation. This result indicates that the ensembling two correlated models cannot lower the prediction errors. For the band gap energy, averaging the $n$-gram+NN (MAE = 124 meV) and SOAP+GPR (MAE = 98 meV) models yields an MAE = 97 meV ($r = 0.67$) which is lower than the 1st place $n$-gram+KRR model (MAE = 114 meV). These results demonstrate that the Pearson correlation allows for an identification models with de-correlated predictions, which can be combined to obtain even lower errors.

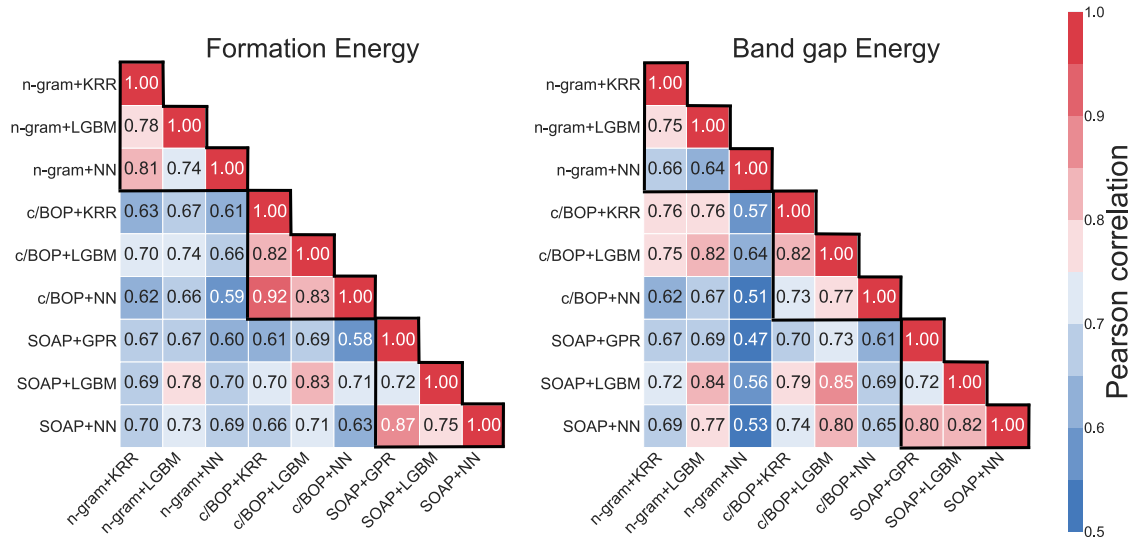

Figure 3. Pearson correlation in the test set errors of the formation energy (left) and bandgap energy (right) between each of the nine combinations of representation and regression model examined in this study. The black triangles indicate the predictions obtained for the same representation.

## III. Training and test set errors using features derived from relaxed structures

For the purposes of efficient predictions in high throughput screening, it is important to incorporate structural features without performing a geometry optimization. If atomic structural information were required from optimized geometries, then most other quantities would be known as well and no predictions were necessary. The discussion has so far been limited to a dataset constructed using geometries generated from the weighted average of the optimized pure binary crystalline systems (i.e., applying Vegard's law[58,59] to generate the input structures). However, the target formation and band gap energies correspond to the fully optimized structures with the lattice vectors and atomic positions allowed to relax self-consistently. Therefore, to examine the additional challenge for the ML description using this structure generation procedure, the performance of the top three ML approaches using the fully relaxed geometries is also examined.

A difference of 3, 1, and 12 meV/cation in the formation energy is calculated between training the $n$-gram+KRR, c/BOP+LGBM, and SOAP+NN approaches using features generated from the relaxed structures compared with the Vegard's law starting structures (Table 2). A similar trend is observed for the bandgap energy where a difference of 7, 7, and 21 meV, respectively, between the predictions using the two sets of geometries. The small difference in the error between the $n$-gram model for the relaxed geometry is attributed to the to the rigid definition of the coordination numbers based on pre-determined cutoff value based on the ionic radii for the bond distances considered within the coordination sphere. In the $n$-gram model, the parameterization of the coordination environment for each lattice symmetry augments the additional challenge of the Vegard's law starting structure by inputting bias into the model; however, this then leads to a representation that is less flexible to different input structures. In contrast, the SOAP representation is strongly dependent on the geometry used for building the descriptor, which leads to a large difference in errors between the two structures.

Table 2. Comparison of test set MAE values for the different regression methods re-trained using fully relaxed geometries for the NOMAD 2018 Kaggle dataset compared with idealized geometries.

| Representation | Regression method | Band gap energy | | Formation energy | |
|---|---|---|---|---|---|
| | | Mean absolute error (meV) | | Mean absolute error (meV/cation) | |
| $n$-gram | KRR | 113 | 106[*] | 17 | 14[*] |
| c/BOP | LGBM | 100 | 93[*] | 14 | 15[*] |
| SOAP | NN | 77 | 98[*] | 1 | 13[*] |

[*] Calculated using features from the Vegard's law starting structure.

## IV. Examining the model generalizability to lattices outside of the training set

Each model was re-trained on a dataset that contained only five out of six lattice structures and then tested on a dataset containing only the lattice structure excluded from the training set. The $Ia\bar{3}$ lattice was chosen as the test set in this investigation because it displays the largest difference in the bandgap minimum and maximum values of all of the lattices (4.42 eV) with a standard deviation of 0.99 eV. The model performance for this re-partitioned training set (2384 structures encompassing five lattice symmetries) and the test set (616 structures of the $Ia\bar{3}$ symmetry) results in significantly larger MAE values of 53 meV/cation, 40 meV/cation, and 110 meV/cation for the formation energy for $n$-gram+KRR, c/BOP+LGBM, and SOAP+NN, respectively (Table S3). A similar increase in the band gap energies is also observed for $n$-gram+KRR (MAE = 179 meV), c/BOP+LGBM (MAE = 180 meV), and SOAP+NN (MAE = 280 meV), respectively. The significant increase in the errors compared with the original dataset is attributed to the absence of common local atomic environment descriptors between the training and test sets.

To examine if an improved generalizability of each model can be obtained by training a model for each lattice type separately, the c/BOP+LGBM model is re-trained by performing the feature selection and hyperparameter optimization procedure for each spacegroup separately and then tested on the left-out $Ia\bar{3}$ lattice. This procedure results in a test set MAE score of 36 meV/cation (111 meV) for the formation (band gap) energy, which is improved compared to the MAE of 40 meV/cation (180 meV) when training the model to the entire training set.

To give an indication of the prediction quality of these three ML models for the left-out lattice, a CE model was examined using a random training/test 75%/25% split of the 616 structures with the Ia$\bar{3}$ lattice symmetry. Using a CE model that includes two-point clusters up to six angstroms, a test set MAE of 23 meV/cation for the formation energy is obtained. A saturation in the learning curve with a training set size of only 50 samples for the CE approach (Figure S3), which indicates that this approach is incapable of achieving a higher accuracy with more data. In comparison to the CE test-set accuracy, the *n*-gram+KRR, and c/BOP+LGBM have about twice the error when the Ia$\bar{3}$ lattice is completely left out of the training set. For the band gap energy, the *n*-gram+KRR and c/BOP+LGBM models are much more accurate compared to what is achieved with CE (229 meV). These results indicate that the simple CE representation still provides a competitive accuracy on for modeling formation energy for lattices left out of the training set; however, the disadvantage of the CE approach is that a new model would have to be trained for each symmetry because only lattice-site occupations are included in this method.

## Discussion

We have presented the three top performing machine learning models for the prediction of two key properties of transparent conducting oxides during a public crowd-sourced data-analytics competition organized by NOMAD and hosted by the online platform Kaggle. One key outcome of this competition was the development of a new representation for materials science based on the *n*-gram model. Because of the diverse set of methods and regression techniques, the interplay between the combination of the representation and regression methods was also analyzed. In particular, consistently large differences between the mean absolute errors and the 95 percentile distributions of the training and test set errors are consistently observed when a neural network and light gradient boosting machine is used at the regression models, which indicates a higher potential for overfitting for these methods. The Pearson correlation was used to investigate correlations between the estimates of the test set values among the various ML models to give additional insight into the model performance. Using this analysis, the largest Pearson correlations were observed for predictions from the same representations combined with different regressors for the formation energy. In particular, the highest predictions were observed for the same representations using neural network and kernel ridge regression (Gaussian process regression). The Pearson correlation allows for an identification models with de-correlated predictions to obtain even lower errors through ensembling.

# Methods

## I. *n*-gram model

The 1st place winning solution uses a crystal graph representation to convert the crystalline structures into features by counting the contiguous sequences of unique atomic sites of various lengths (called *n*-grams).[60] In this crystal graph representation (see Figure 4), the nodes correspond to an atom in the unit cell and the edges between nodes are defined by the coordination environment. In this approach, the coordination environment of each atom was determined by counting the metal-oxygen distances that are less than the sum of ionic Shannon experimental radii[70] scaled by 130-150% depending on the lattice type. In this crystalline graph generated for the unit cell, a directed graph with parallel edges to account for the periodicity (i.e., a given node that sits on the edge of the unit cell may have additional and equivalent bonds if translational symmetry is applied). Previously a crystal graph representation (constructed using a different definition of the coordination environment) was employed to create a consistent discretized representation of solid-state lattice (e.g., the cubic $ABX_3$ perovskite lattice), which could then be used directly with convolutional neural network for learning properties of materials.[71] Although the *n*-gram model also relies on the discretization of the lattice, features were generated by binning the nodes of the contiguous sequences along a path in the crystal graph (see "Path Graph" in Figure 4) varying from 1 (unigram) to 4 (quadgram).

Several *n*-gram items for a specific path in the crystal graph are labeled in Figure 4. The unigram features are generated from counting the unique coordination environments present along this labeled path (i.e., two four-coordinate gallium atoms [Ga-4], one two-coordinate oxygen [O-2], two three-coordinate oxygen atoms [O-2 and O-3], one five-coordinate indium [In-5]). The histogram of bigrams is formed from the contiguous sequence of two nodes, which corresponds to the combination of nearest-neighbor nodes (i.e., two nodes that share an edge). Trigrams and quadgrams are contiguous sequences up to three nodes and four nodes, respectively. In the example presented in Figure 4, the complete set of bigrams is two O-3/Ga-4, two O-2/Ga-4, and one O-3/In-5 bigrams.

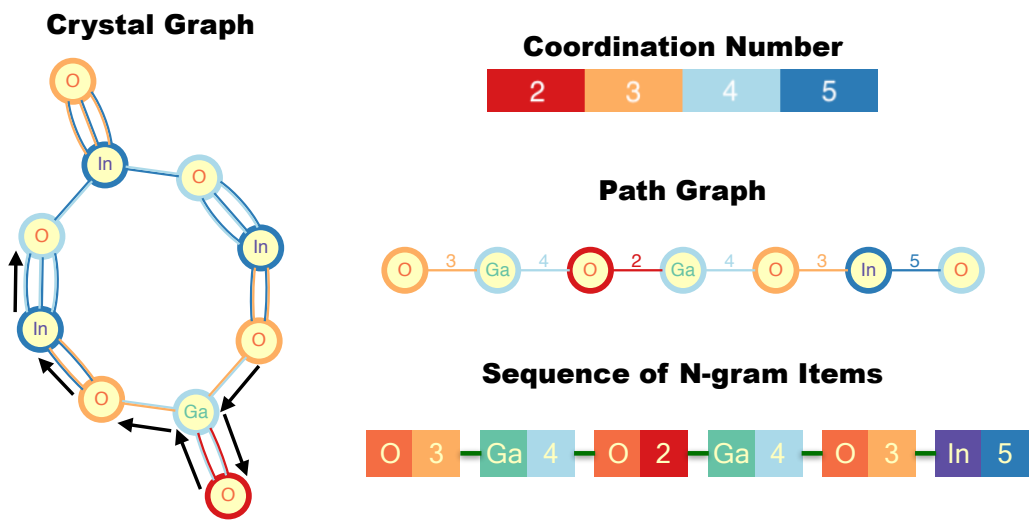

Figure 4. Depiction of crystal graph representing a configuration of $In_3Ga_1O_6$, which shows the connections between each node that are defined by the chemical bonds.

For the NOMAD 2018 Kaggle dataset, a total of 13 unique unigrams were used that range from 4 to 6 and unique oxygen coordination numbers that range from 2 to 5. To illustrate the histogram features generated from the *n*-gram model using, the unigrams for two 80-atom structures with the formula $(Al_{0.25}Ga_{0.28}In_{0.47})_2O_3$ and $(Al_{0.63}Ga_{0.34}In_{0.03})_2O_3$ and C/2m and P6$_3$/*mmc* symmetry types are shown in Figure 5. Because of the variation in the count of *n*-grams for structure with different unit-cell sizes, these features were therefore normalized by the unit cell volume.

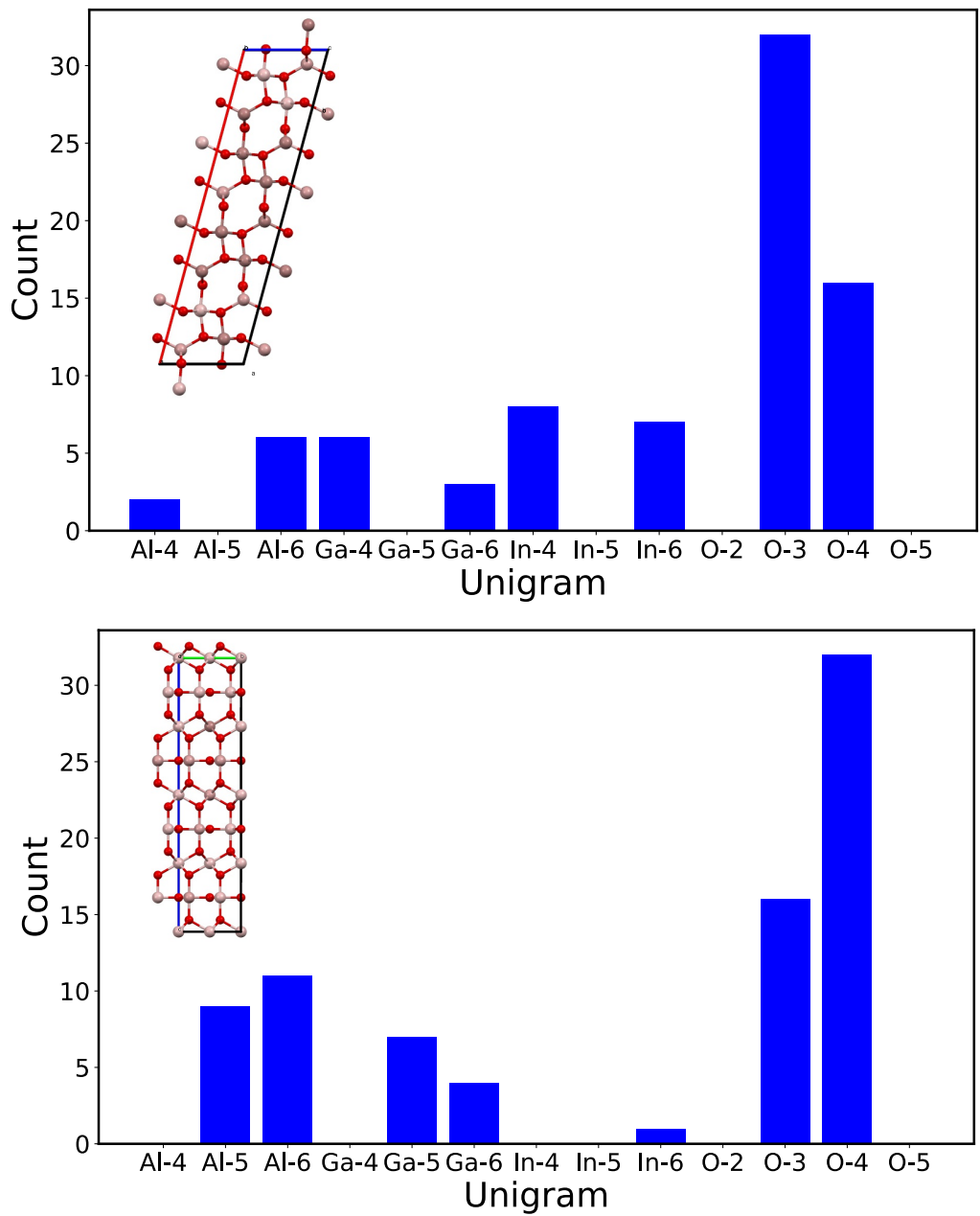

Figure 5. Histogram of the complete set of 13 unigram features formed from the total list of the unique coordination environment for each atom type for two training-set structures $(Al_{0.25}Ga_{0.28}In_{0.47})_2O_3$ and $(Al_{0.63}Ga_{0.34}In_{0.03})_2O_3$.

The $n$-gram features are combined with a KRR model using the Gaussian radial basis function kernel. The values of the two hyperparameters ($\alpha_i$, the weights of each sample $i$ and γ is the length scale of the Gaussian, which controls the degree of correlation between training point) were determined by performing grid searches with 5-fold CV and compares well to the private leaderboard score (Table 3). Similar to what was discussed in the context of ensembling different models with low correlation, here too the highest accuracies are obtained from an ensemble score of the trigram and quadgram predictions: $P_{\text{mix}} = a_{\text{mix}}P(\text{trigram}) + (1 - a_{\text{mix}})P(\text{quadgram})$, where a mixing parameter of 0.64 and 0.69 for the formation and band gap energies was used, respectively. Although such an ensemble gives the lowest RMSLE, the entire list of unigrams, bigrams, trigrams, and quadgram features, were used throughout the discussion presented in this paper to facilitate a comparison between each of the different regression methods. This is a convenient choice to avoid having to re-train the mixing parameter for each analysis.

Table 3. CV-score of the formation energies and band gap energies and public and private leaderboards RMSLE values for *n*-grams of various lengths (normalized by unit cell volume).

| *n*-grams lengths | Formation energy RMSLE | Bandgap RMSLE | Public RMSLE | Private RMSLE |
| --- | --- | --- | --- | --- |
| Unigram | 0.0229 | 0.0817 | 0.0518 | 0.0560 |
| Bigram | 0.0230 | 0.0811 | 0.0472 | 0.0540 |
| Trigram | 0.0223 | 0.0814 | 0.0381 | 0.0514 |
| Ensemble of trigram and quadgram | 0.0200 | 0.0780 | 0.0380 | 0.0510 |
| Quadgram | 0.0210 | 0.0770 | 0.0394 | 0.0506 |

The *n*-gram features were combined with a NN architecture consisting of 11 dense layers with 100, 50, 50, 20, 20, 20, 20, 10, 10, 10 neurons respectively and LeakyReLU activation functions. The NN was implemented in PyTorch[72] and optimized using Adam[73] with a learning rate of 0.005. The *n*-gram features were combined with LGBM with the model hyperparameter optimization performed as described in Section II of the Methods.

## II. Atomic and Bond-order-potential derived features

For the 2nd place model, many descriptor candidates are examined from a set of compositional, atomic environment-based, and average structural properties (Figure 6). Of this list, the optimal 175 (212) features are selected for the prediction of the band gap (formation) energy are based on an iterative procedure using the auxiliary gradient boosting regression tree (XGBoost) and used with the LGBM learning algorithm.

The weighted chemical properties are computed from reference data using either the overall stoichiometry or the nearest neighbors. This approach is motivated by the concepts of structure maps that chart the structural stability of compounds in terms of chemical properties of the constituent atoms and the overall chemical composition.[74-76] For generating per-structure features, the weighted arithmetic mean of band gap and of formation energy are computed from the stoichiometry using the respective values for $In_2O_3$ (R-3c, Ia3, Pnma), $Ga_2O_3$ (C/2m, R-3c), and $Al_2O_3$ (C/2m, Pna21, R-3c, and $P4_232$) of the Materials Project.[77] The average and difference of several free-atom properties such as the electronic affinity, ionization potential, atomic volume, and covalent radius (all values were obtained from Ref. [78]) are computed between each atom and each of its nearest neighbors to generate per-atom features. The list of nearest atomic neighbors is generated using the ASE package[79] and determined based on the distance between two atoms being smaller than sum of the computed free-atom radii.

The representations of the atomic environment are incorporated using BOP-based properties and simple geometric measures. The latter are comprised of averaged atomic bond distances, averaged cation-oxygen nearest-neighbor bond distances, centrosymmetric parameters (determined from a sum of the vectors formed between atom $i$ and its nearest neighbors); and the volume per atom. The characterization of atomic environments by the BOP methodology relies on moments and the closely related recursion coefficients that connect the local atomic environment and local electronic structure (DOS) by the moments theorem.[80] Within the analytic BOP formulism, these properties can be computed efficiently in an approximate way[83,61,63] and used as per-atom features that represent the local atomic environment.[64,81] For each atom, the *n*-th moment is computed by multiplying pairwise model Hamiltonians along self-returning paths (*i.e.*, start and end at the same atom) up to length *n*. BOP allow for the discrimination and classification of atomic structures[81] and local atomic environments,[64] and therefore, make possible structural properties based on the atomic environment. In this work, a total of 12 moments corresponding to the atomic environment up to the 6th nearest neighbor shell was used. This procedure is to some degree comparable to the *n*-gram approach of the 1$^{st}$ place solution with regard to sampling the environment. For example, a quadgram would correspond to one half of a self-returning path in an 8$^{th}$ moment calculation. One of the differences in the two methodologies is that all path segments are used explicitly in the *n*-gram approach whereas only the individual self-returning paths are subsumed in the moments of the c/BOP approach.

For each atom in the structure, this procedure generates a list with a length that is equivalent to the number of neighbors. A clustering scheme is then applied to the average and standard deviation of these features is used to generate a fixed-length representation. These properties were clustered into seven effective-atom groups based on its atomic environment described by $a_1^{(j)}$, $b_2^{(j)}$, and $v_j$ using the k-means clustering algorithm[75,76] applied separately to O and Al, Ga, and In for each structure in the dataset. These clusters of varying lengths were then projected into a fixed-length vector by taking only the mean and standard deviation. If one of the 7 effective atoms is not present in a given structure, then the corresponding feature is set to zero.

In total, this approach resulted in a set of 6,950 features (*ca.* 120 atomic properties per atom × 7 effective atomic environments × 4 element types × 2 statistical aggregation measures), which were reduced to set of 175 and 212 features for the prediction of the band gap and formation energies that produced the highest accuracy based on an iterative procedure using XGBoost.[82] The final set of features where then combined with LGBM[53] for the final model with the hyper-parameters tuned using 10-fold CV within the hyperopt package[83] and a suggestion algorithm using tree-structured Parzen estimators,[84] which resulted in an RMSLE value of 0.0462 and 0.0521 for the public and private leaderboards.

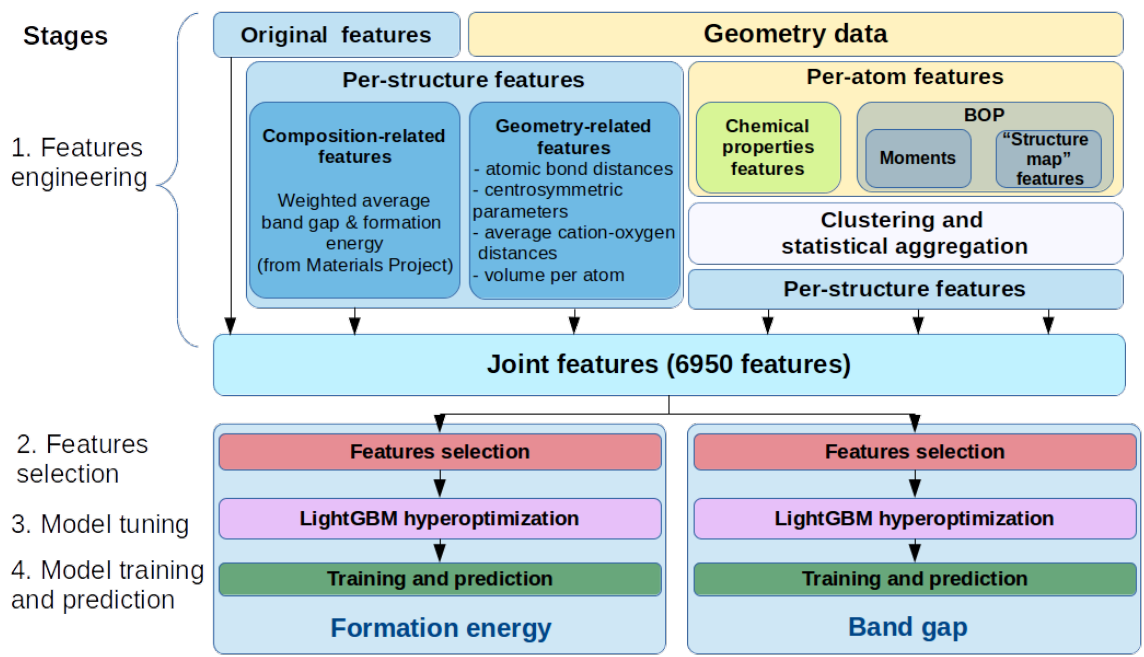

Figure 6. Illustration of feature engineering and subsequent stages for the construction of the 2nd place c/BOP descriptor.

The selection of the optimal set of features requires attributing an importance to each of ca. 7,000 features. However, recently, popular feature attribution methods were shown to have a lower assigned importance relative the true impact of that feature in modeling the target property.[85] The SHapley Additive exPlanations (SHAP) method[86] was proposed to give more accurate relative features importances and were calculated here as a normalized mean absolute value of the SHAP values for each feature (see Figure S4). For prediction of the band gap energy, the features with the largest relative importance (*ca.* 17% each) are the weighted band gap of $Al_2O_3$, $Ga_2O_3$, and $In_2O_3$ and the volume per atom. In contrast, all features have a relatively small importance for the prediction of the formation energy; only geometrical information describing the environment of indium and the length centrosymmetric parameter has the highest importance. The per-atom features have a total relative importance of 40% and 33% for formation energy and band gap energy, respectively, including *ca.* 20% and 15% of the relative feature importance for the BOP-related features

The same set of top-features used with LGBM to achieve the 2$^{nd}$ place score were also combined with KRR and NN. The features used with the KRR and NN regressors were rescaled to have a zero mean and unit variance. The KRR model employed a Gaussian radial basis function kernel with the α and γ hyperparameters tuned using a 5-fold CV grid search. The Keras package[87] with the Tensorflow backend[88] was used to generate a three-layer NN containing 1,024, 256 and 256 neurons with batch normalization, hyperbolic tangent activation function and 20% dropout in each layer. The output layer contained one neuron only had no batch normalization and used an ReLU activation function.[89] The NN were trained for 500 epochs.

## III. SOAP feature vector

The 3$^{rd}$ place solution used the smooth overlap of atomic positions (SOAP) kernel developed by Bartók et al. that incorporates information on the local atomic environment through a rotationally integrated overlap of Gaussian densities of the neighboring atoms.[13,14] The SOAP kernel describes the local environment for a given atom through the sum of Gaussians centered on each of the atomic neighbors within a specific cutoff radius. The SOAP vector was computed using the QUIPPY package[90] using a real-space radial cutoff in $f_{cut}$ of 10 Å and the smoothing parameter $\sigma_{atom}$ = 0.5 Å. The basis set expansion values of $l$ = 4 and $n$ = 4 were also used. For each structure, a single feature vector was used by averaging the per-atom SOAP vector for each atom in the unit cell, which resulted in a vector with a length of 681 values. These aggregated mean feature vectors for the dataset were then scaled so that each dimension has a mean equal to zero and variance equal to one.

The average SOAP features were used in a three-layer feed-forward NN using Pytorch with batch normalization and 20% dropout in each layer. For predicting the bandgap energies and the formation energies, the initial layer had 1024 neurons and 512 neurons, respectively. In both cases, the remaining two layers had 256 neurons each. The neural networks were trained for 200 and 250 epochs for the prediction of the bandgap energies and the formation energies, respectively. The final predictions were based on 200 independently trained NNs using the same architecture but with different initial weights.

The average SOAP vector of each structure was combined with Gaussian Process Regression (GPR),[18] where the covariance function between two structures was defined as a polynomial kernel:

$$k(R_i, R_j) = (aR_i \cdot R_j + b)^c$$

Where $R_i$ and $R_j$ are descriptor vectors for structure $i$ and $j$; $a$, $b$, and $c$ are kernel coefficients. The SOAP kernel can be re-written as:

$$K(R_i, R_j) = \left( \sum_{n_1 n_2 l} P_{n_1 n_2 l}(R_i) P_{n_1 n_2 l}(R_j) \right)^x$$

Several values for the Polynomial kernel degree $x$ (ranging from 1 – 6) with $a = 1.0$ and $b = 0.0$ were examined until the lowest RMSLE was obtained. This resulted in two hyperparameters for the model construction: regularization of the GPR and the degree of the kernel. Optimal hyperparameters were identified using repeated random sub-sampling CV for 100 training and validation splits. The predictive accuracy was assessed using the validation data. Finally, the final GPR model was averaged over all 100 splits, which resulted in optimal regularization values of $7.6 \times 10^{-6}$ and $3.84 \times 10^{-5}$ for the formation energy and bandgap energy, respectively. These settings resulted in a RMSLE of 0.021 and 0.085 for the formation energy and band gap energy for the test set.

The SOAP vector was also combined with LGBM regression with the model hyperparameter optimization performed as described in Section II. This combination has proven to be suboptimal (discussed in the main text) and was dropped in favor of using a NN in the final submission.

## Acknowledgments

The idea of organizing a competition was given to us by Bernhard Schölkopf and Samuel Kaski, who are on the scientific advisory committee of the NOMAD. We also gratefully acknowledge the help of Will Cukierski from Kaggle for assistance in launching the competition. The NOMAD Kaggle competition award committee members are Claudia Draxl, Daan Frenkel, Kristian Thygesen, Samuel Kaski, and Bernhard Schölkopf. In addition, we thank Gabor Csanyi, Matthias Rupp, Mario Boley, Christopher Bartel, and Marcel Langer for the helpful discussions and providing feedback on this manuscript. The project received funding from the European Union's Horizon 2020 research and innovation program (grant agreement no. 676580) and the Molecular Simulations from First Principles (MS1P). C.S. gratefully acknowledges funding by the Alexander von Humboldt Foundation. A single-point calculation using the Vegard's law starting geometry for each structure in the Kaggle competition is available in the NOMAD repository.

## Additional Information

The supporting information contains the training and test sets of the competition. We also provided: a distribution of the formation energies and bandgap energies for the training and test set; tabulated results from randomly re-partitioning the entire dataset; learning curves for each of the models; the associated information with Fig. 2; tabulated results from re-partitioning the dataset to exclude one lattice symmetry in training set; results from the cluster expansion model; and a summary of the features selected by the $2^{nd}$ place model of this competition.

## Competing Interests

The authors declare no competing interests.

# Supporting Information for:
# NOMAD 2018 Kaggle Competition: Solving Materials Science Challenges Through Crowd Sourcing

Christopher Sutton,[1,*] Luca M. Ghiringhelli,[1]
Takenori Yamamoto,[2] Yury Lysogorskiy,[3] Lars Blumenthal,[4,5]
Thomas Hammerschmidt,[3] Jacek Golebiowski,[4,5] Xiangyue Liu,[1] Angelo Ziletti,[1]
Matthias Scheffler[1,*]

[1]*Fritz Haber Institute of the Max Planck Society*
*Berlin, Germany*

[2] *Research Institute for Mathematical and Computational Sciences (RIMCS), LLC*
*Yokohama, Japan*

[3] *ICAMS*
*Ruhr-Universität*
*Bochum, Germany*

[4] *EPSRC Centre for Doctoral Training on Theory and Simulation of Materials*
*Department of Physics*
*Imperial College London*
*London, U.K.*

[5] *Thomas Young Centre for Theory and Simulation of Materials*
*Department of Materials*
*Imperial College London*
*London, U.K.*

[*]Corresponding authors: sutton@fhi-berlin.mpg.de; ghiringhelli@fhi-berlin.mpg.de

The two target properties of the NOMAD 2018 Kaggle competition were the formation energy and band gap energy. The formation energy is calculated relative to pure $In_2O_3$, $Al_2O_3$, and $Ga_2O_3$ phases and were normalized per number of cations according to:

$$E_f = E[(Al_xGa_yIn_z)_2O_3] - xE[Al_2O_3] - yE[Ga_2O_3] - zE[In_2O_3]$$

where $x, y,$ and $z$ are the corresponding relative concentrations of Al, Ga, and In, respectively defined as: $x = \frac{N_{Al}}{N_{Al}+N_{Ga}+N_{In}}, y = \frac{N_{Ga}}{N_{Al}+N_{Ga}+N_{In}}, z = \frac{N_{In}}{N_{Al}+N_{Ga}+N_{In}}$. $E[(Al_xGa_yIn_z)_2O_3]$ is the energy of the mixed system, $E[Al_2O_3]$, $E[Ga_2O_3]$, and $E[In_2O_3]$ are the energies of the pure binary crystalline systems in their thermodynamic ground state. This relative formation energy provides an estimate of the stability of the mixed system with respect to the stable ground state of the binary components and differs from the usual definition that instead uses the atomic energies for reference values. The relative formation energy is instead used because the use of atomic energies incorporates a large linear trend into the definition of the formation energy, which is easily learned by any machine learning approach and results in much lower errors for the formation energy compared to the bandgap energy. By using the bulk energies of the pure binary components as reference values (instead of the atomic energies), a more similar in in these two target properties is obtained. A similar distribution of the two target properties between the training and test sets for the formation energy (top, Figure S1) and bandgap energy (bottom, Figure S1) are observed for the 2400-value training and 600-value test set used in the NOMAD 2018 Kaggle competition.

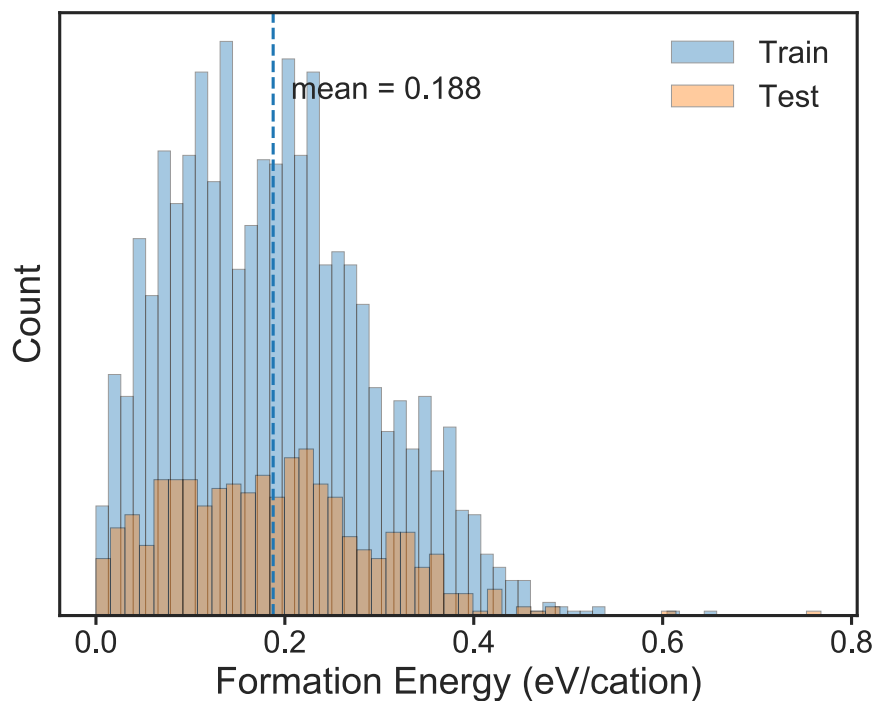

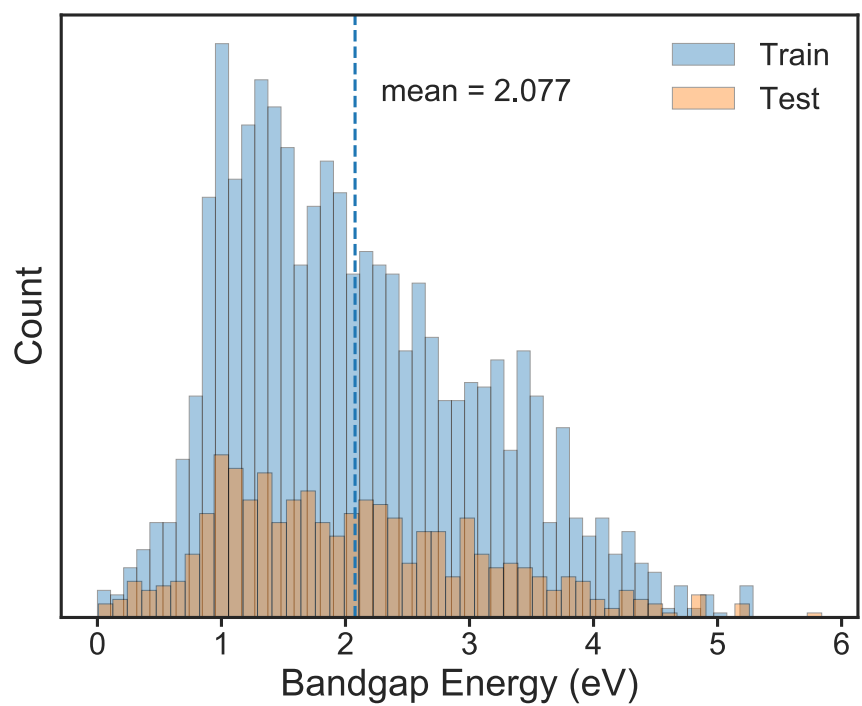

Figure S1. Histogram of the formation energy (top) and band gap energy (bottom) for the training and test sets used in the NOMAD 2018 Kaggle competition.

Table S1. Comparison of the average and standard diversion of MAE values for the formation and bandgap energies for five random 80%/20% training/test set splits with the MAE values for the dataset used in the NOMAD 2018 Kaggle competition (Original).

| Descriptor | Method | Formation energy (meV/cation) | | Band gap energy (meV) | |
| --- | --- | --- | --- | --- | --- |
| | | Avg. (Std.) five 80/20 splits | Original | Avg. (Std.) five 80/20 splits | Original |
| *n*-gram | KRR | 16 (0.6) | 14 | 119 (5) | 106 |
| atomic/BOP | LGBM | 17 (0.1) | 15 | 104 (2) | 93 |
| SOAP | NN | 14 (0.1) | 13 | 107 (9) | 99 |

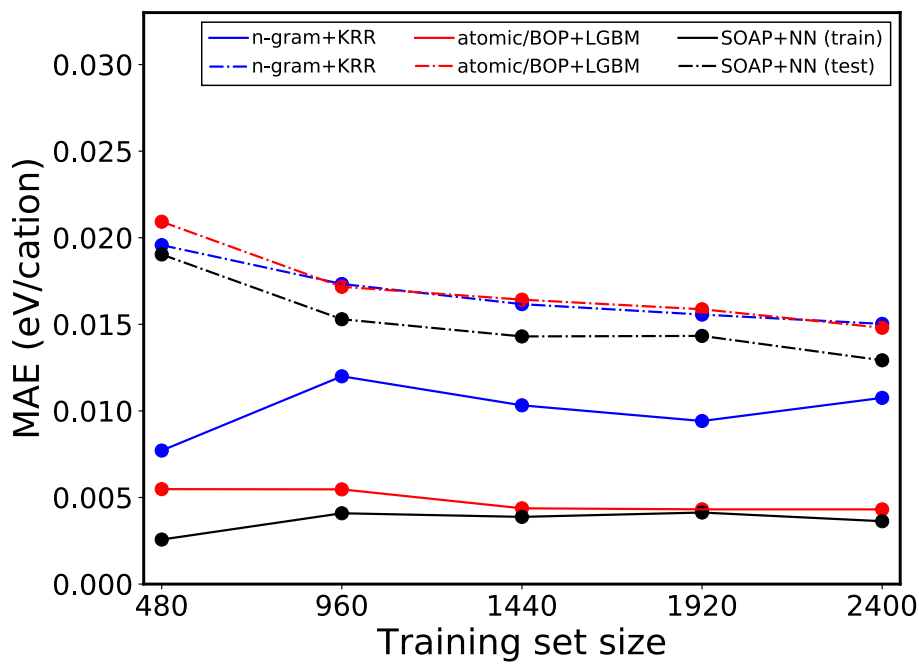

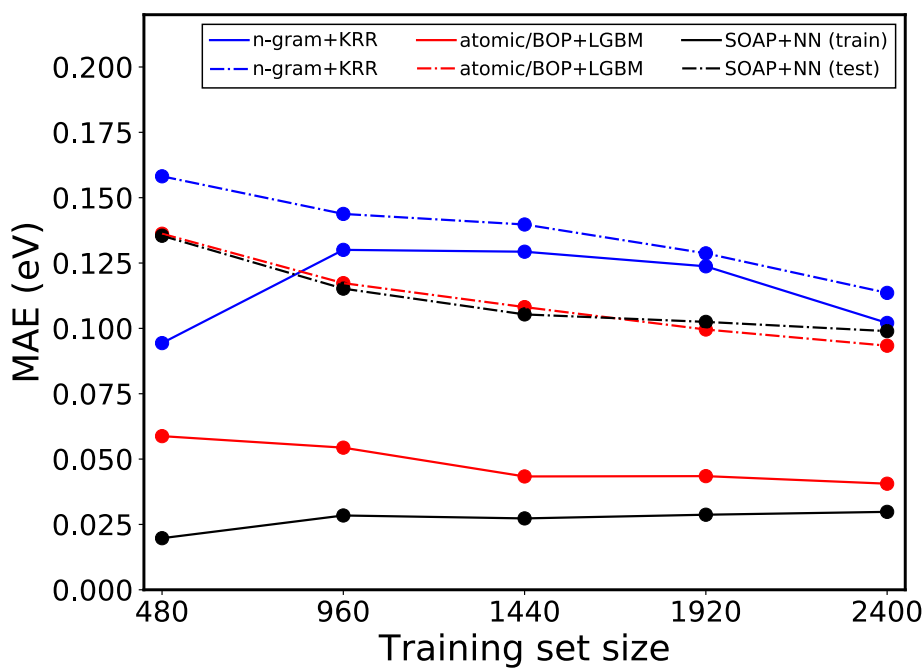

Figure S2. MAE learning curves for the formation energy (top) and bandgap (bottom) the training (solid lines) and test set (dashed) for *n*-gram+KRR (blue), atomic/BOP+LGBM (red), and SOAP+NN (black) using random subsets in increments of 20% of the 2400-samples training with a consistent 600-sample test set that was used in the NOMAD 2018 Kaggle competition (*i.e.*, 480 samples in the training set, with predictions made on the same 600-sample test set), 40% (960/600), 60% (1440/600), 80% (1920/600) and 100% (2400/600). All three models were trained using 5-fold cross-validation.

Table S2. A comparison of the three winning representations of the competition combined with the KRR/GPR, NN, and LGBM regression models. The mean absolute error (MAE) of the formation energy and band gap energy for the test set and their 95% confidence values for the MAE are also provided.

| Representation | Regression model | Formation energy (meV/cation) | | Band gap energy (meV) | |
|---|---|---|---|---|---|
| | | Training (95% per.) | Test (95% per.) | Training (95% per.) | Test (95% per.) |
| $n$-gram | NN | 6 (18) | 16 (54) | 65 (188) | 124 (397) |
| $n$-gram | KRR | 11 (36) | 14 (55) | 96 (337) | 106 (352) |
| $n$-gram | LGBM | 4 (12) | 17 (61) | 39 (129) | 110 (399) |
| c/BOP | NN | 13 (36) | 19 (64) | 77 (229) | 118 (410) |
| c/BOP | KRR | 9 (27) | 17 (61) | 93 (319) | 118 (452) |
| c/BOP | LGBM | 4 (12) | 15 (51) | 41 (135) | 94 (393) |
| SOAP | NN | 4 (10) | 13 (46) | 32 (83) | 99 (392) |
| SOAP | GPR | 5 (16) | 13 (47) | 35 (107) | 98 (378) |
| SOAP | LGBM | 2 (6) | 15 (60) | 23 (92) | 110 (478) |

Table S3. Comparison of test set MAE values for the three winning models trained for a dataset containing five of the total six lattice symmetries and a test set on comprising of only one lattice symmetry (Ia$\bar{3}$) contained in the full 3000-sample (Al$_x$Ga$_y$In$_z$)$_2$O$_3$ dataset.

| Representation | Regressor | Formation energy (meV/cation) | Band gap energy (meV) |
|---|---|---|---|
| $n$-gram | KRR | 53 | 179 |
| SOAP | NN (GPR) | 11 (11) | 280 (680) |
| atomic/BOP-features | LGBM | 40[*]/36[**] | 180[*]/111[**] |

* Feature selection and model hyper-optimization according to five-fold cross-validation with splits generated randomly

** Feature selection and model hyper-optimization according to five-fold cross-validation with splits generated based on spacegroup number

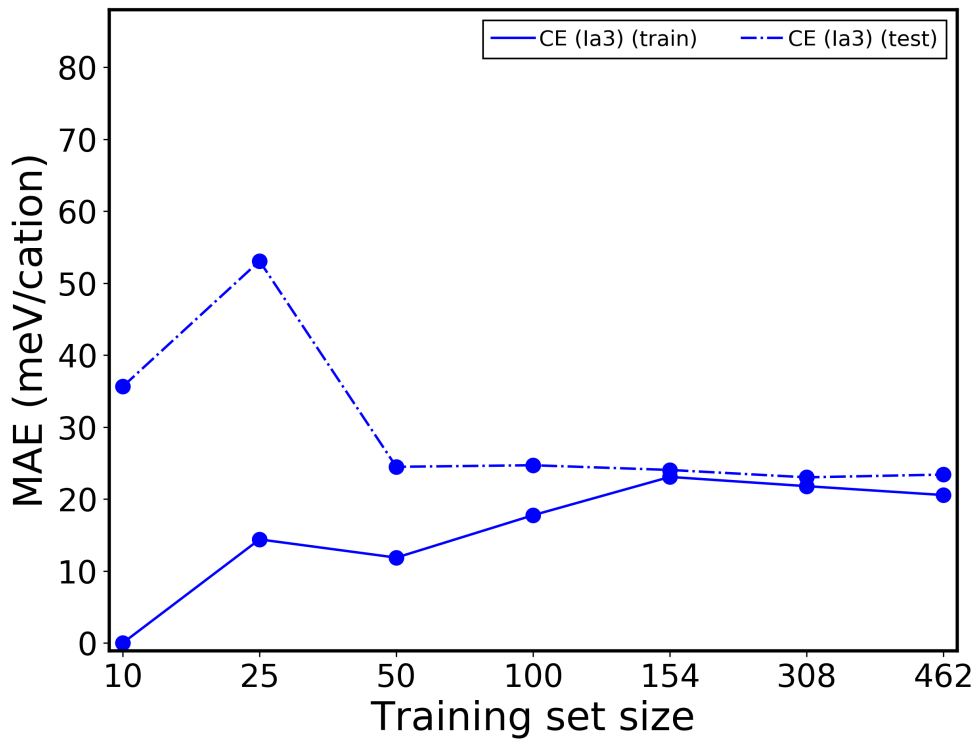

Figure S3. MAE learning curves for the formation energy (top) for the training (solid lines) and test set (dashed) using a cluster expansion model constructed from a set of two-point clusters up to six-angstroms.

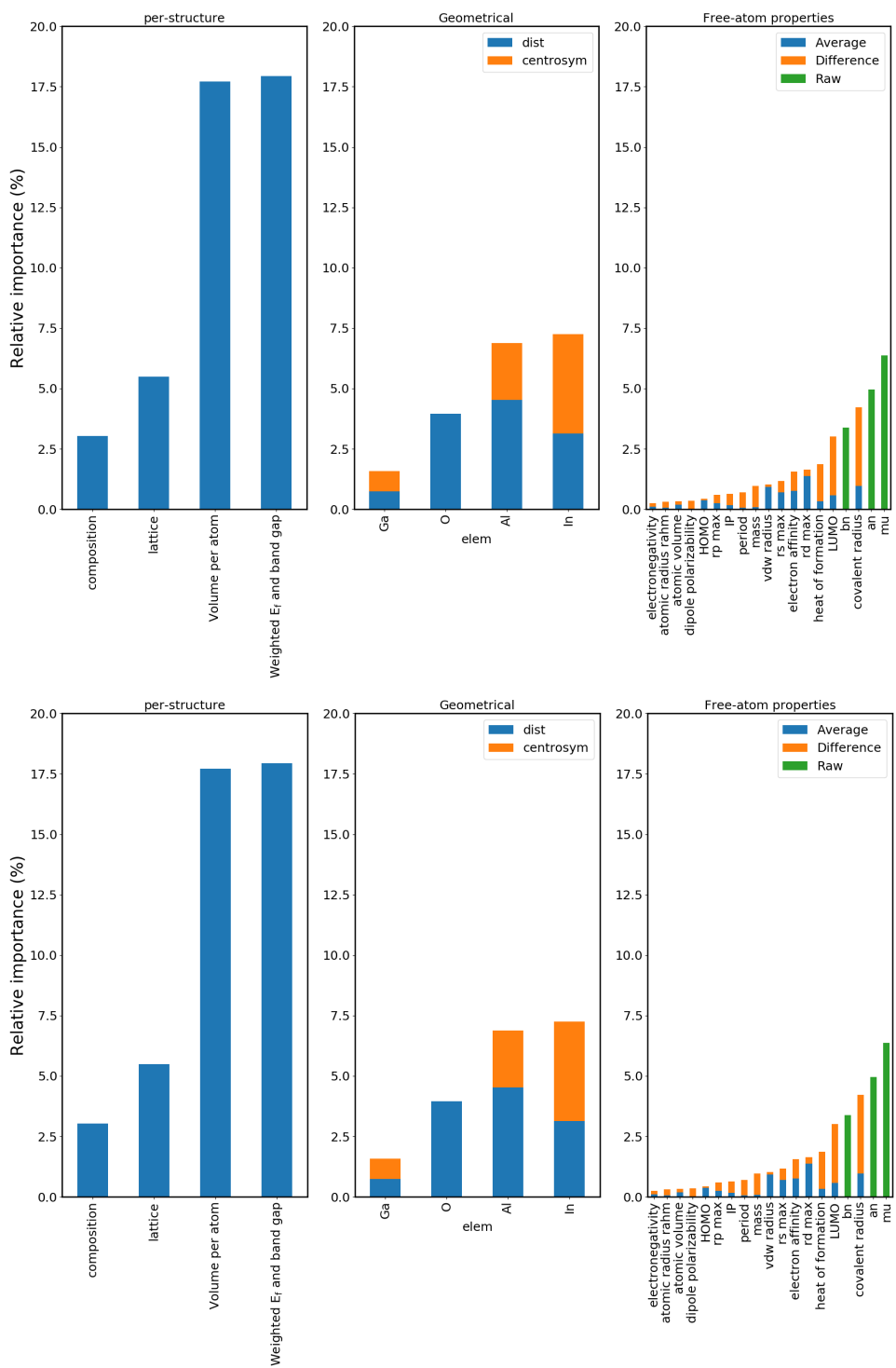

Figure S4. Relative importances for different groups of features for the formation energy (top) and band gap energy (bottom) per-structural features are comprised of compositional-related features and lattice-vector lengths.